# Concentration dependence of the fluorescence decay profile in transition metal doped chalcogenide glass

M. Hughes[a], D.W. Hewak[a] and R.J. Curry[b],
[a]Optoelectronics Research Centre, University of Southampton, Southampton, SO17 1BJ, United Kingdom
[b]Advanced Technology Institute, School of Electronics and Physical Sciences, University of Surrey, Guildford, GU2 7XH, United Kingdom

## Abstract

In this paper we present the fluorescence decay profiles of vanadium and titanium doped gallium lanthanum sulphide (GLS) glass at various doping concentrations between 0.01 and 1% (molar). We demonstrate that below a critical doping concentration the fluorescence decay profile can be fitted with the stretched exponential function: $\exp[-(t/\tau)^{\beta}]$, where τ is the fluorescence lifetime and β is the stretch factor. At low concentrations the lifetime for vanadium and titanium doped GLS was 30 μs and 67 μs respectively. We validate the use of the stretched exponential model and discuss the possible microscopic phenomenon it arises from. We also demonstrate that above a critical doping concentration of around 0.1% (molar) the fluorescence decay profile can be fitted with the double exponential function: $a*\exp-(t/\tau_1)+ b*\exp-(t/\tau_2)$, where $\tau_1$and $\tau_2$ are characteristic fast and slow components of the fluorescence decay profile, for vanadium the fast and slow components are 5 μs and 30 μs respectively and for titanium they are 15 μs and 67 μs respectively. We also show that the fluorescence lifetime of vanadium and titanium at low concentrations in the oxide rich host gallium lanthanum oxy-sulphide (GLSO) is 43 μs and 97 μs respectively, which is longer than that in GLS. From this we deduce that vanadium and titanium fluorescing ions preferentially substitute into high efficiency oxide sites until at a critical concentration they become saturated and low efficiency sulphide sites start to be filled.

## 1. INTRODUCTION

Many relaxation processes in complex condensed systems such as glasses have long been known to follow the Kohlraush-Williams-Watts (KWW) function, which is also currently know as the stretched exponential function and is given in equation 1

$$I(t) = I_0 \exp\left(-\left(\frac{t}{\tau}\right)^{\beta}\right) \tag{1}$$

Where τ is a characteristic relaxation time and β is the stretch factor ranging between 0 and 1. The closer β is to 0 the more the function deviates from a single exponential. Stretched exponential relaxation was first described by Kohlraush in 1847 to model the decay of the residual charge on a glass Leyden jar. Since then the stretched exponential function has been shown to fit many other relaxation processes in amorphous materials such as nuclear relaxation,[1] magnetic susceptibility relaxation,[1] fluorescence decay[2] and photoinduced dichroism.[3] Stretched exponential behaviour is rarely observed in crystalline solids; with some exceptions.[4] There have been several mutually exclusive microscopic explanations for the observed stretched exponential relaxation in glasses, this can be see as being due to the use of different models of the fundamental structure of glasses. The models for stretched exponential relaxation behaviour in glasses generally fall into two categories: spatially heterogeneous dynamics and temporally heterogeneous dynamics. The spatially heterogeneous dynamic model assumes that the relaxation of a single excited ion follows a single exponential law and that the system remains homogeneous over the time taken for the relaxation to occur. Validation of this model has been allowed through the development of fluorescence microscopy which allows the study of single molecules in complex condensed environments[5]. It has been shown that that the fluorescence decay of a single molecule in varying local environments leads to a stretched-exponential decay as a result of the presence of a continuous distribution of lifetimes[6,7]. The temporally heterogeneous dynamics model assumes that the system remains homogeneous in space but random sinks, in disordered material such as glass, capture excitations and become progressively depleted with time. This causes the decay rate itself to slow with the progress of time, stretching the decay. These two processes may occur at the same time in a given relaxation process.

## 2. EXPERIMENTAL AND ANALYSIS

Samples of vanadium and titanium doped gallium lanthanum sulphide (GLS) were prepared by mixing 65% gallium sulphide, 30-x% lanthanum sulphide, 5% lanthanum oxide and the appropriate proportion (x%) of vanadium or titanium sulphide (% molar). Samples of vanadium and titanium doped gallium lanthanum oxy-sulphide (GLSO) were prepared by mixing 75-x% gallium sulphide, 25% lanthanum oxide and the appropriate proportion (x%) of vanadium or titanium sulphide (% molar). The melt components were mixed in a dry-nitrogen purged glove box. Gallium and lanthanum sulphides were synthesised in-house from gallium metal (9N purity) and lanthanum fluoride (5N purity) precursors in a flowing $H_2S$ gas system. Before sulphurisation lanthanum fluoride was purified and dehydrated in a dry-argon purged furnace at 1250 $^oC$ for 36 hours to reduce $OH^-$ and transition metal impurities. The lanthanum oxide and vanadium sulphide were purchased commercially and used without further purification. The glass was melted in a dry-argon purged furnace at 1150 $^oC$ for 24 hours before being quenched and annealed at 400 ºC for 12 hours.

Fluorescence lifetime measurements where obtained by exciting the vanadium doped GLS samples with a 1064 nm Nd:YAG laser and the titanium doped GLS samples with a 658 nm laser diode. The excitation sources were modulated by an acousto-optic modulator (AOM) and attenuated to give an incident power on the sample of around 10mW in order to minimise heating. The fluorescence was detected with a New Focus 2053 InGaS detector which was set to a gain that corresponded to a 3dB bandwidth of 3MHz. The data was captured by a Picosope ADC-212 virtual oscilloscope with a 12 bit intensity resolution and a maximum temporal resolution of 700 ns, the signal was averaged for around 2 minutes to improve signal to noise ratio in the measurement. Lifetime measurements where taken several times at different alignments, different time frames on the oscilloscope and different detector frequency responses in order to give an estimate of random experimental error.

Regression analysis was implemented using the Marquardt-Levenberg algorithm,[8] given in equation 2. This algorithm seeks the values of the parameters that minimize the sum of the squared (SS) differences between the values of the observed and predicted values of the dependent variable.

$$SS = \sum_{i=1}^{n} (y_i - \hat{y}_i)^2 \qquad (2)$$

Where $y_i$ is the observed and $\hat{y}_i$ is the predicted value of the dependent variable, the index i refers to the $i^{th}$ data point and n is the total number of data points. The goodness of the fit was measured using the coefficient of determination ($R^2$), given in equation 3.

$$R^2 = \frac{\sum_{i=1}^{n} (\hat{y}_i - \bar{y})^2}{\sum_{i=1}^{n} (y_i - \bar{y})^2} \qquad (3)$$

Where $\bar{y}$ is the mean value of the observed dependant variable. The coefficient of determination indicates how much of the total variation in the dependent variable can be accounted for by the regressor function. If $R^2 = 1$ this indicates that the fitted model explains all variability in the observed dependant variable, while $R^2 = 0$ indicates no linear relationship between the dependant variable and regressors.

## 3. RESULTS AND DISCUSSION

Figure 1 shows the fluorescence decay of 0.09% vanadium doped GLS (V:GLS) fitted with a stretched exponential. The best fit to the experimental data was with a lifetime of 30 μs and a stretch factor (β) of 0.8. Visual inspection indicates an excellent fit to the experimental data, this is confirmed with an $R^2$ of 0.9996.

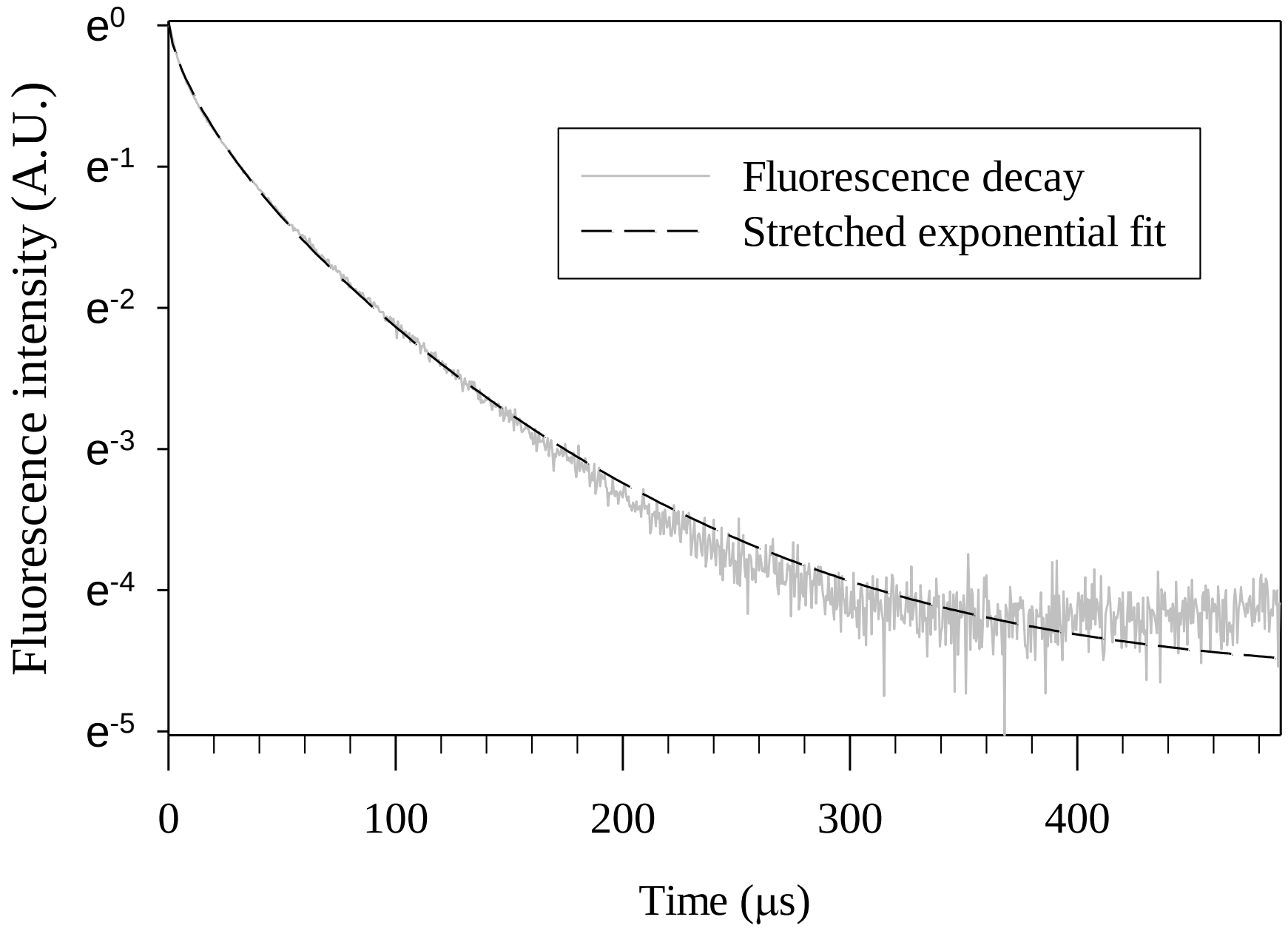

Fig.1 Fluorescence decay of 0.09% vanadium doped GLS fitted with a stretched exponential. The lifetime was 30 μs and the stretch factor was 0.8

Figure 2 shows the fluorescence decay of 1% V:GLS fitted with a stretched and double exponential function. The double exponential function is given in equation 4.

$$I(t) = I_1 \exp - \left( \frac{t}{\tau_1} \right) + I_2 \exp - \left( \frac{t}{\tau_2} \right) \quad (4)$$

Where $\tau_1$ and $\tau_2$ are the two characteristic lifetimes; $I_1$ and $I_2$ are their respective coefficients. Inspection reveals the stretched exponential does not describe the data as well as at lower concentrations and the double exponential function is a better fit. The $R^2$ for the double and stretched functions are 0.9945 and 0.9839 respectively. The lifetimes for the double exponential fit are 5 and 29 μs for both 0.5 and 1% V:GLS; this is significant as it indicates that the lifetime observed at low concentrations is still present at high concentrations. It is noted that since we are modelling the decay in our glass with the stretched exponential model we should fit this decay with a double stretched exponential. However in order to minimised the number of regressor variables so that a fair comparison can be made between the two models a double exponential is thought to be a good approximation. This is because a single exponential function can be fitted to a stretched exponential (β = 0.8) with $R^2$ = 0.9839.

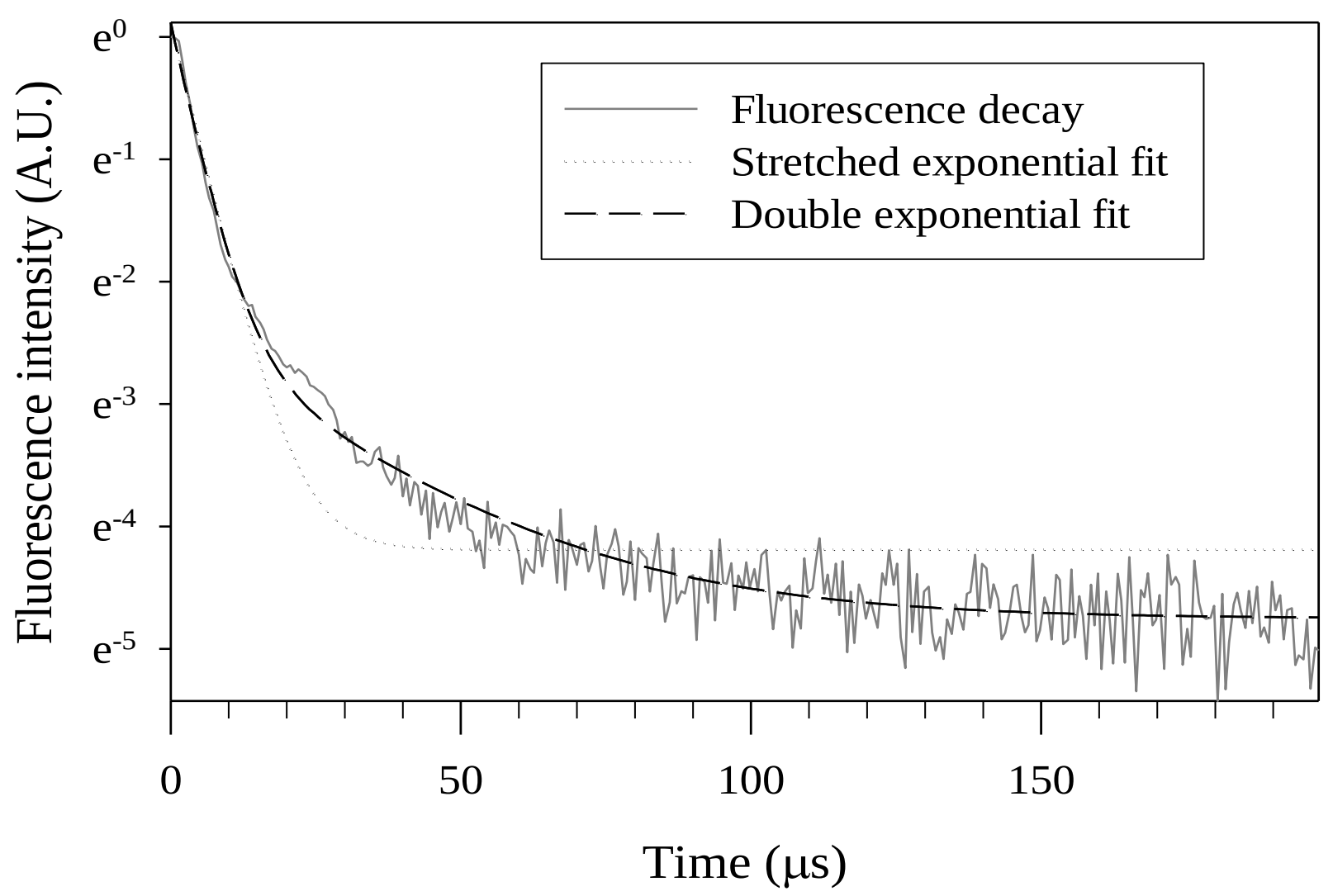

Fig. 2 Fluorescence decay of 1% vanadium doped GLS fitted with a stretched and a double exponential $\tau_1$ =29 μs $\tau_2$ = 5 μs

Figure 3 shows the $R^2$ of stretched and double exponential fits as a function of vanadium concentration. Figure 3 indicates the stretched exponential gives an almost perfect fit for vanadium concentrations up to 0.1%. The double exponential also gives a good fit at these concentrations but a double exponential can give an almost perfect for to an artificially generated stretched exponential decay. So as long as the decay can be accurately described by a stretched exponential we will assume it is the correct model for the decay. Above ~0.1% concentration the fluorescence decay starts to deviate from stretched exponential behaviour and can be more accurately described by a double exponential. The lifetime of V:GLS was ~ 30 μs for concentrations between 0.01 and 0.1% this indicates that the observed stretched exponential behaviour, at least for these concentrations, was entirely due to a continuous distribution of lifetimes. This is because if it were caused by progressively depleted random sinks one would expect the lifetime to decrease with increasing concentration. The fluorescence decay of 0.01 % V:GLSO was found to have a lifetime of 43 μs and was a good fit (R2 = 0.9997) to a stretched exponential. Higher doping concentrations of V:GLSO have yet to be fabricated.

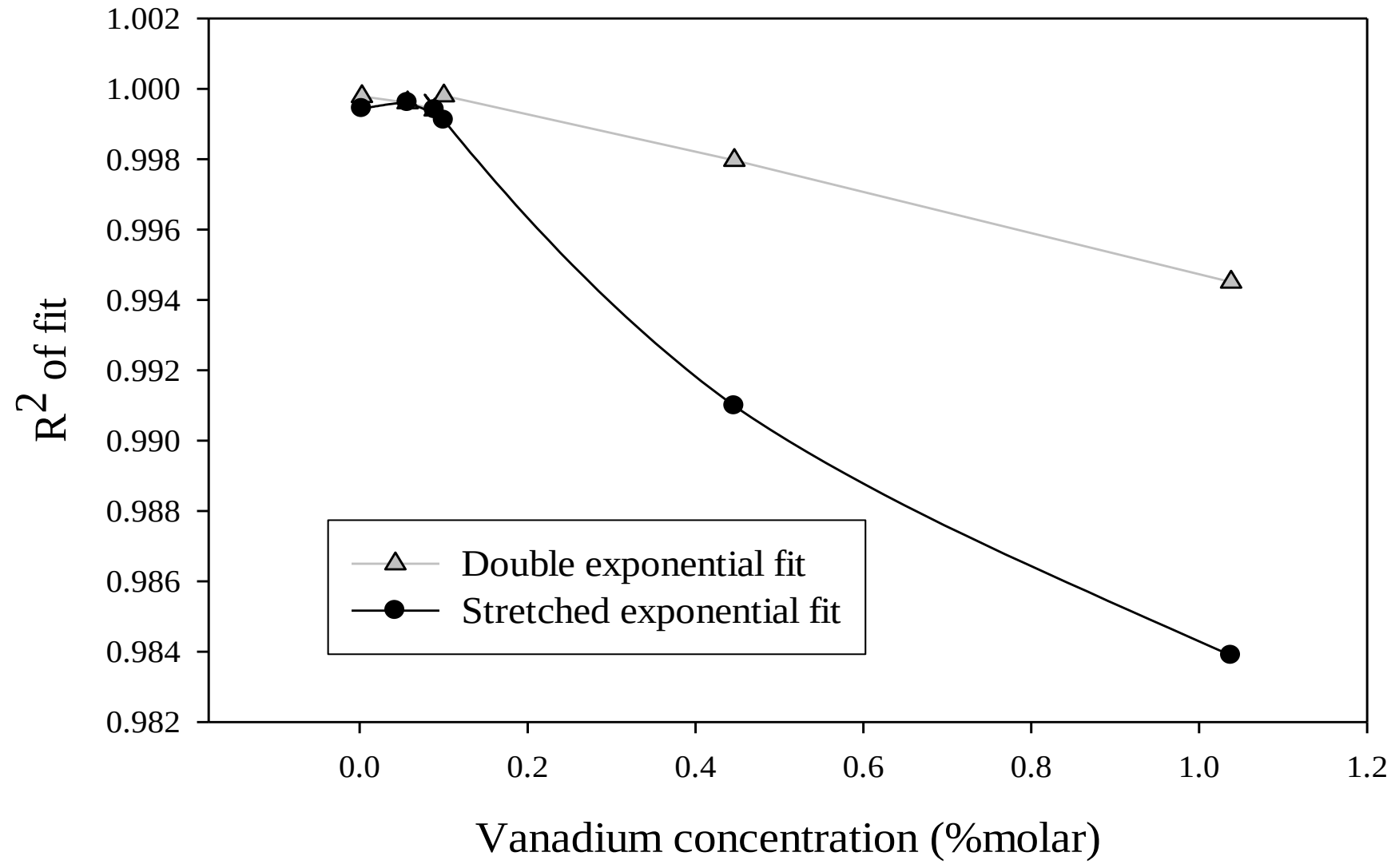

Fig. 3 $R^2$ of stretched and double exponential fits as a function of vanadium concentration. Lines are a guide for the eye

Figure 4 shows the fluorescence decay of 0.05% titanium doped GLS (Ti:GLS) fitted with a stretched exponential. A lifetime of 67 µs and a stretch factor was 0.5 were fitted. Similarly to low concentration V:GLS there is a good fit (R2 = 0.9877) to a stretched exponential.

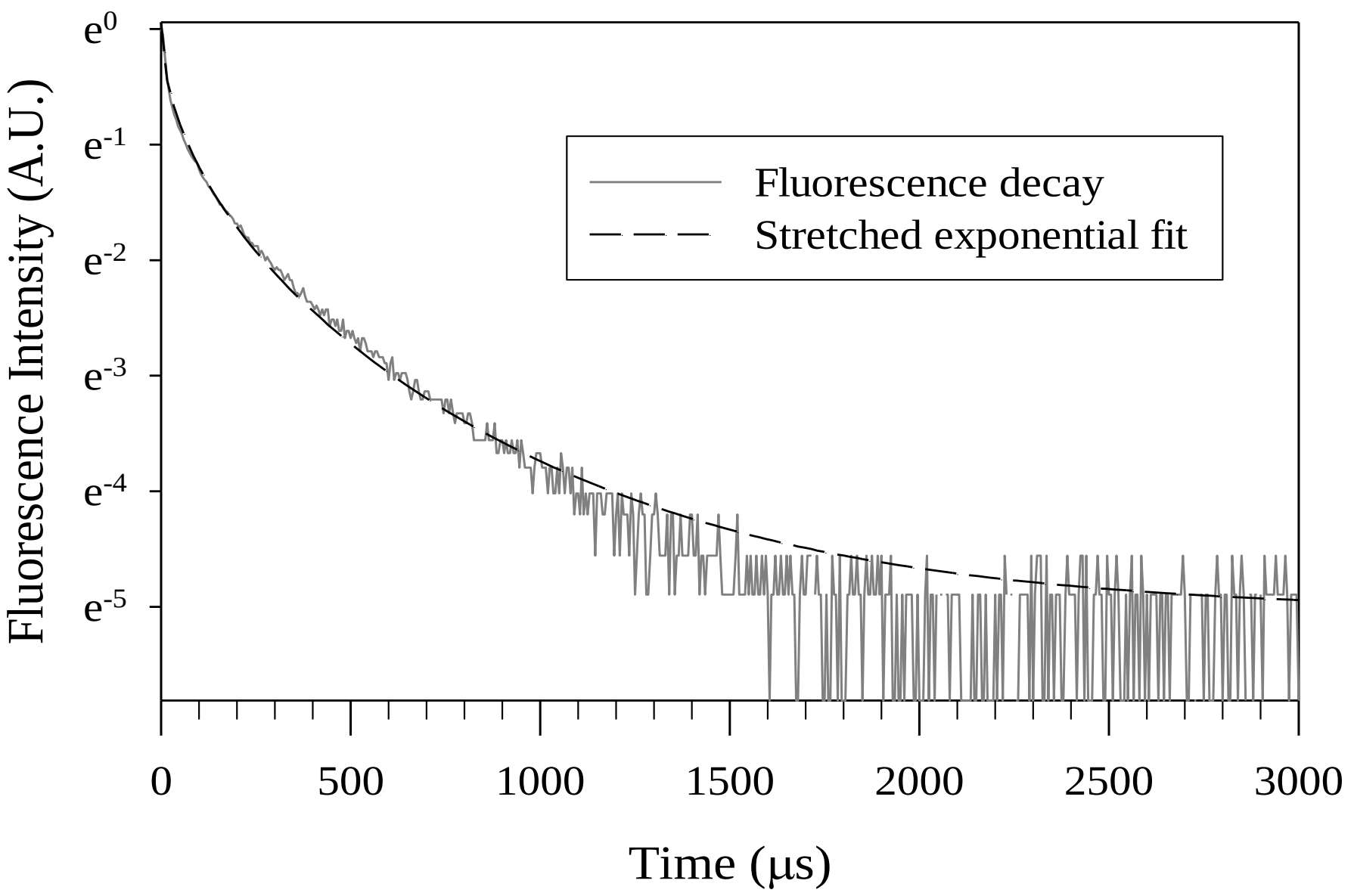

Fig. 4 Fluorescence decay of 0.05% titanium doped GLS fitted with a stretched exponential. The lifetime was 67 µs and the stretch factor was 0.5

The change in stretch factor from 0.8 for V:GLS to 0.5 for Ti:GLS deserves comment. While we do not believe the stretch factor its self can be directly related to a physical phenomenon a comparison between these two dopants is valid. The smaller stretch factor for Ti:GLS indicated that there is a larger distribution of lifetimes than in V:GLS. Since both ions are in the same host the Ti ion must be more strongly influenced by the varying local environments in GLS. The vanadium ion is in a 3+ oxidation state and is tetrahedrally coordinated[9]. Ti:GLS displays one main absorption band centred at 600 nm which indicates that it is in a 3+ oxidation state and has a higher crystal field strength than vanadium. Analysis of various complexes of $V^{3+}$ and $Ti^{3+}$ also indicate that $Ti^{3+}$ would be subject to a higher crystal field strength than $V^{3+}$[10]. This phenomenon can therefore explain the increased stretching for the Ti dopant since ligands of the glass host will be closer to the optically active d-orbitals of the Ti ion and will therefore be more strongly influenced by site to site variations in the crystal field which will hence produce a greater spread of lifetimes.

Figure 5 shows the fluorescence decay of 1% Ti:GLS fitted with the stretched and double exponential. Similarly to 1% V:GLS the stretched exponential is no longer a good fit ($R^2$ = 0.9355) and the double exponential fit is better ($R^2$ = 0.9824). The lifetimes of the double exponential were 15 µs and 160 µs. The fact that the characteristic slow lifetime of the 67 µs is no longer observed at high concentrations when using the double exponential fit as it is in V:GLS is believed to be because  the stretch factor is now 0.5 and a single exponential is no longer a good approximation. Fitting a single exponential to a stretched exponential with a stretch factor of 0.5 gives R2 = 0.8632. To overcome this problem fluorescence intensity data from 0 to 100 µs was discarded, leaving just the slow component of the decay. This was fitted with a stretched exponential with β fixed at 0.5. This fit had the characteristic slow lifetime of ~67 µs. The same procedure for data 0 to 100 µs gave a lifetime of 15 µs.

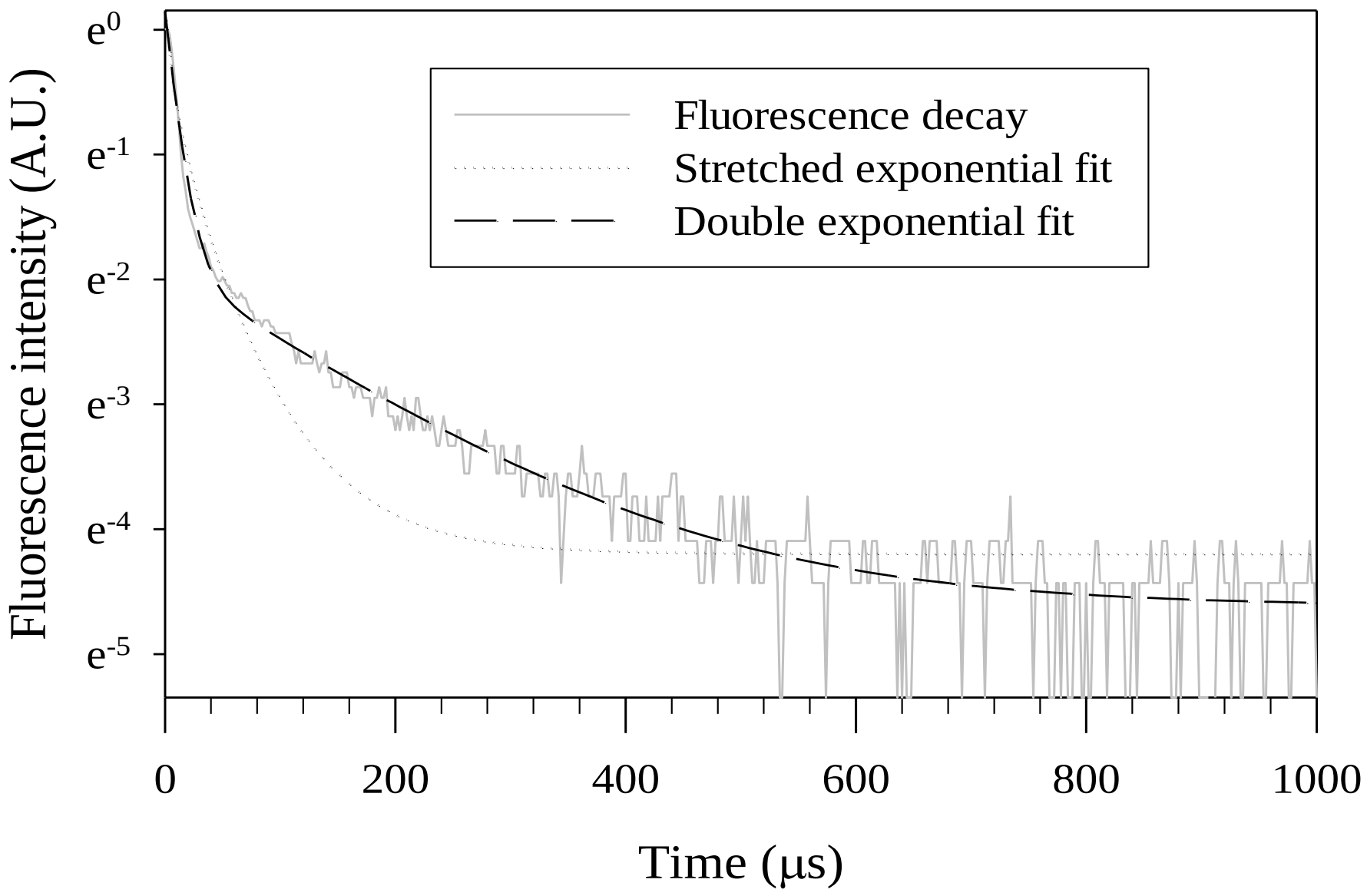

Fig. 5 Fluorescence decay of 1% titanium doped GLS fitted with a stretched and a double exponential

Figures 6 and 7 show the fluorescence decay for 0.05 and 1 % Ti:GLSO fitted with a stretched exponential. Unlike V and Ti:GLS both decays for 0.05 and 1% are well described by the stretched exponential with $R^2$ = 0.9968 and 0.9930 respectively. Figure 8 shows how $R^2$ of a stretched exponential fit varies as a function of titanium concentration in GLS and GLSO. Similarly to V:GLS $R^2$ remains high up to around 0.1% where it starts to fall. In comparison $R^2$ for Ti:GLSO changes very little with Ti concentration. The lifetime for Ti:GLSO drops from 97 to 60 μs at concentration increases from 0.05 to 1% this is probably caused by some form of concentration quenching.

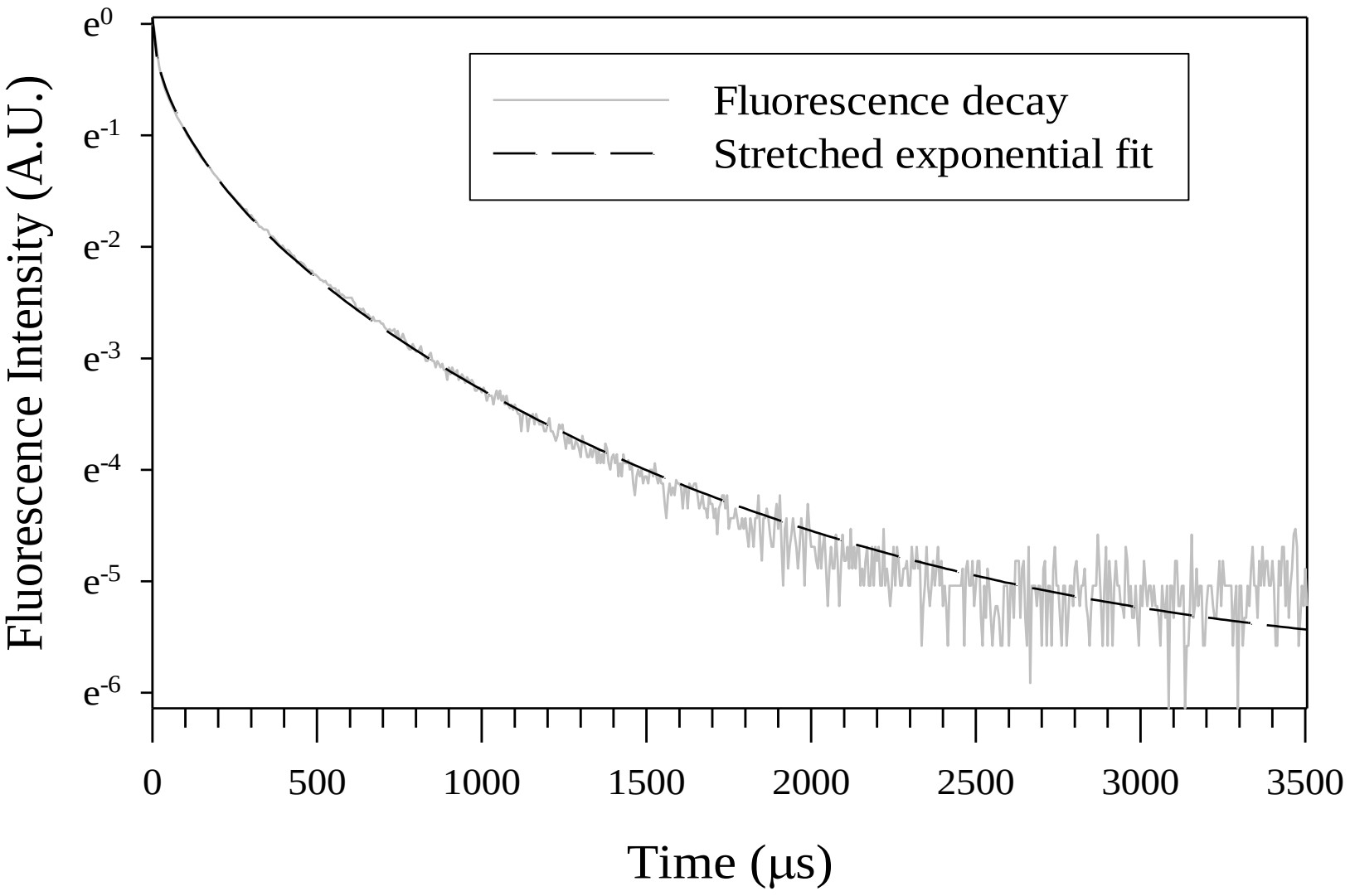

Fig. 6 Fluorescence decay of 0.05% titanium doped GLS0 fitted with a stretched exponential. The lifetime was 97 μs and the stretch factor was 0.53

The decrease in lifetime for Ti:GLS and V:GLS is probably caused, at least in part, by concentration quenching as well. However the fact that there is a deviation from stretched exponential behaviour in concentrations above 0.1% in GLS but not in GLSO and that lifetimes are longer in GLSO indicates that another effect is taking place and that it is something to do with the oxygen content of the glass. GLS contains ~0.5% (molar) oxygen whereas GLSO contains ~15% (molar) oxygen. We therefore propose that two reception sites for transition metals exist in GLS glass; a high efficiency oxide site and a low efficiency sulphide site. In GLS the transition metal ion preferentially fills the high efficiency oxide sites until at a concentration of ~0.1% they become saturated and the low efficiency sulphide sites starts to be filled; this explains the deviation from stretched exponential behaviour at concentrations > 0.1% and the appearance of characteristic fast and slow lifetime components. The peak position and shape of the absorption bands of V and Ti:GLS do not change significantly as concentration increases from 0.01 to 1%, however one would expect a noticeable red shift in absorption going from an oxide coordinated transition metal to a sulphide coordinated transition metal. So the oxide site probably doesn't have oxygen directly bonded to the transition metal ion.

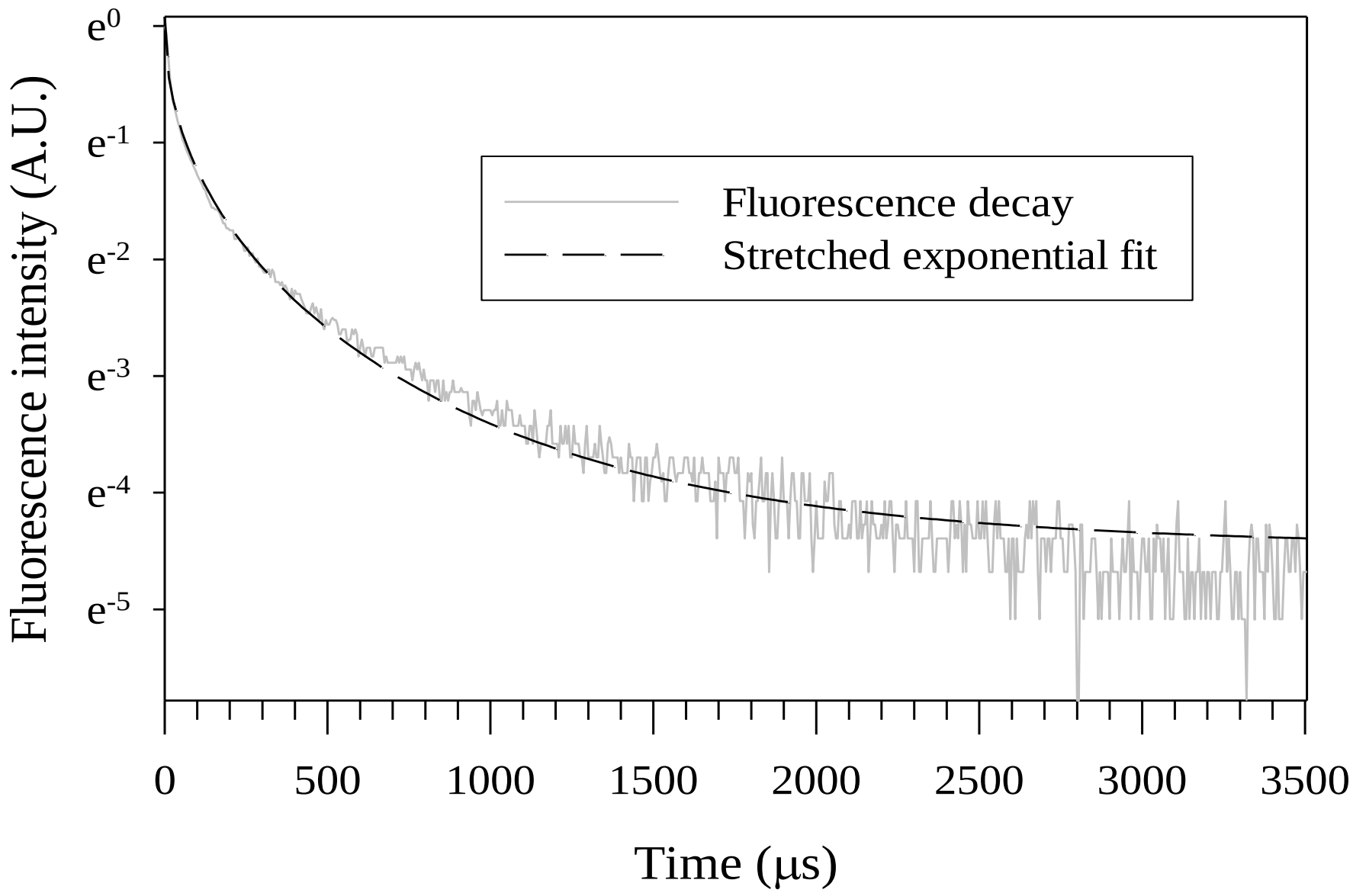

Fig. 7 Fluorescence decay of 1% titanium doped GLS0 fitted with a stretched exponential. the lifetime was 60 μs and the stretch factor was 0.51

To put these findings in a structural context other authors have proposed GLS consists of a covalent network of $GaS_4$ tetrahedra, inter-dispersed by essentially ionic La-S channels.[11-13] In addition some $GaS_4$ tetrahedra with a negative charge are formed from the Ga being bonded to one sulphur anion ($S^{2-}$) to produce "sulphide negative cavities". Oxygen anions ($O^{2-}$) in the glass will similarly produce negatively charged $GaS_3O$ tetrahedra or "oxide negative cavities". These negative ionic cavities form some reception sites for $La^{3+}$ ions, which act as charge compensators for these negative charges.[12,14] Dopant ions in glasses are generally expected to enter substitutionally for network modifier cations[15]. The main network modifier in GLS is $La^{3+}$[12] which is 8 fold coordinated to sulphur with an undetermined symmetry[11]. We therefore propose that V and Ti substitute for $La^{3+}$ and are sulphide coordinated; however in the high efficiency oxide sites one or more of the sulphur atoms is part of an oxide negative cavity whereas in the low efficiency sulphide sites none, one or more of these sulphur atoms is part of a sulphide negative cavity. Two dopant reception sites have been reported for dysprosium doped GLS (Dy:GLS),[14] key differences with this work are that the lifetimes in Dy:GLSO and Dy:GLS were 100 μs and 2.54 ms respectively. This meant that the oxide site was low efficiency and the sulphide site was high efficiency. Separate absorption bands were also identified for the two sites.

Dysprosium has the same outer electron structure as La; so if Dy substitutes for La the 8-fold coordination and any symmetry that may exist would be expected to be maintained. However as a consequence of the outer electron structure of V and Ti, tetrahedral and octahedral coordination are the most likely coordinations to occur. Crystal field calculations and comparisons of spectroscopic data for V:GLS to other work indicates that V:GLS has a 3+ oxidation state and is tetrahedrally coordinated[9]. So the addition of a transition metal to GLS will change the coordination of the La site it substitutes for from 8 to 6 or 4 (4 in the case of V:GLS) and bring about symmetry that may no have already existed. The oxide site being high efficiency for V:GLS and Ti:GLS, but low efficiency for Dy:GLS can be explained because O is more electronegative than S so the oxide site will have a slightly higher crystal field strength than the sulphide site. The separation of the lowest energy levels in transition metals is strongly influenced by crystal field strength so there will be a greater separation of the two lowest energy levels and therefore a lower probability of non-radiative decay. Energy levels in rare earths on the other hand are influenced very little by crystal field strength. The oxide site would be expected to have a higher phonon energy than a sulphide site[14] this would increase the probability of non-radiative decay for both transition metal and rare-earth dopants. If the decrease in non-radiative decay caused by the increase in crystal field strength is greater than the increase in non-radiative decay caused by the increase in phonon energy for a transition metal in an oxide site then this would explain why the oxide site is high efficiency for transition metals and low efficiency for rare-earths.

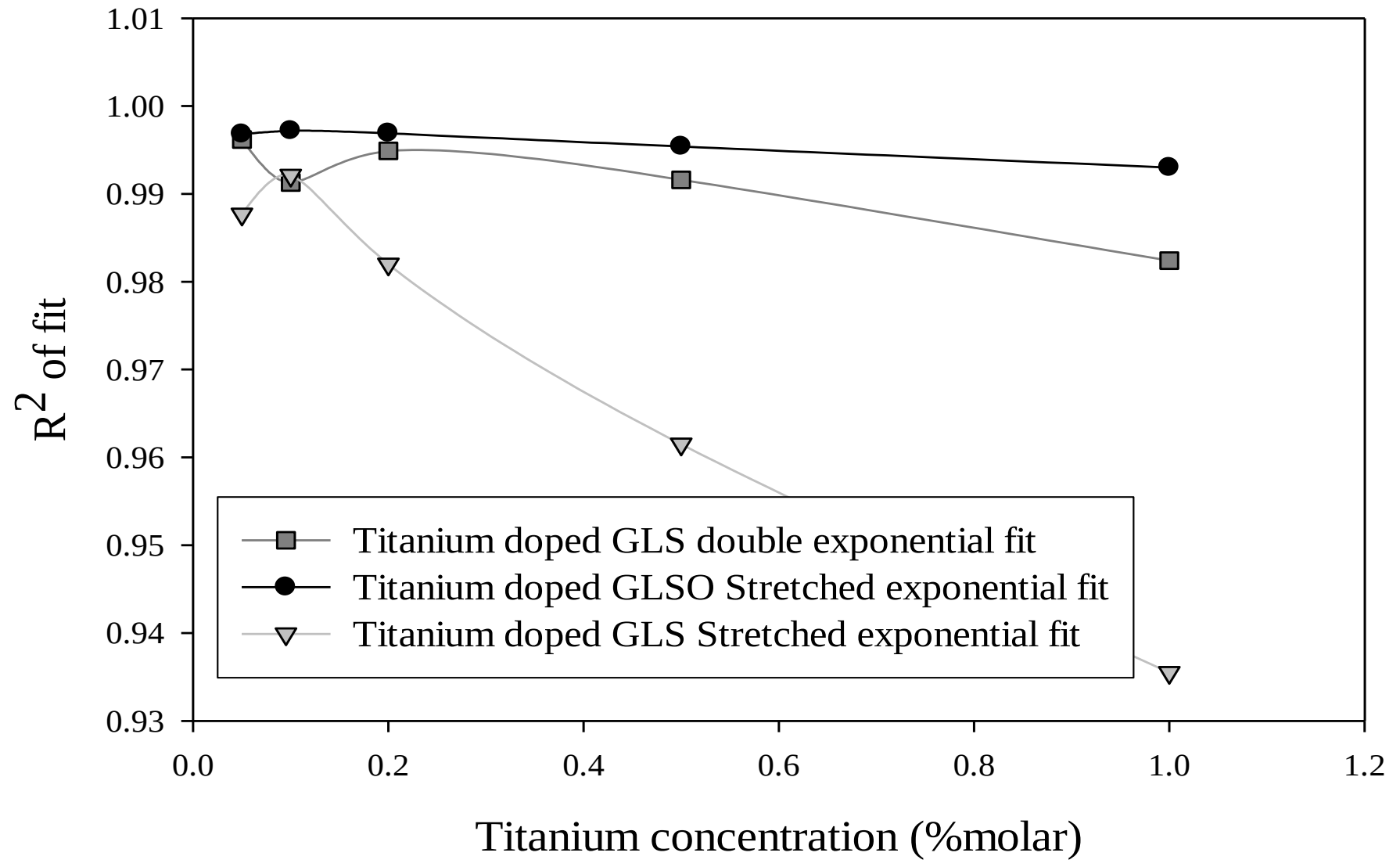

Fig. 8 of stretched and double exponential fits as a function of vanadium concentration

## 4. CONCLUSIONS

By analysing the goodness of fit of stretched exponentials to the fluorescence decay of vanadium and titanium doped GLS and GLSO at various doping concentrations we have show that two reception sites exist for the dopants. These are a high efficiency oxide site that gives rise to characteristic slow lifetimes and a low efficiency sulphide site that gives rise to characteristic short lifetimes. We give an explanation why in dysprosium doped GLS the oxide site is low efficiency and the sulphide site is high efficiency. We relate the decrease in stretch factor from 0.8 for V:GLS to 0.5 for Ti:GLS to optically active electrons in V:GLS being further away from the host ligands in Ti:GLS because of the higher crystal field strength in Ti:GLS.